\documentclass[english,jcp,preprint,endfloats,superscriptaddress]{revtex4-1} 
\usepackage{epsfig} 
\usepackage{amsmath} 
\usepackage{graphicx} 
\usepackage{longtable} 
\usepackage{hyperref} 
\usepackage{amssymb} 
\usepackage{dcolumn} 
\usepackage{mhchem} 
\usepackage{subfigure} 
\usepackage{array} 
\usepackage{multirow} 
\usepackage{tabularx} 
\usepackage{cleveref} 
\begin{document} 

\title{SCAN-based hybrid and double-hybrid density functionals from models without fitted parameters} 

\author{Kerwin Hui} 
\affiliation{Department of Physics, National Taiwan University, Taipei 10617, Taiwan} 

\author{Jeng-Da Chai} 
\email[Author to whom correspondence should be addressed. Electronic mail: ]{jdchai@phys.ntu.edu.tw} 
\affiliation{Department of Physics, National Taiwan University, Taipei 10617, Taiwan} 
\affiliation{Center for Theoretical Sciences and Center for Quantum Science and Engineering, National Taiwan University, Taipei 10617, Taiwan} 

\date{\today} 

\begin{abstract} 

By incorporating the nonempirical SCAN semilocal density functional [Sun, Ruzsinszky, and Perdew, Phys. Rev. Lett. {\bf 115}, 036402 (2015)] in the underlying expression of four existing hybrid and 
double-hybrid models, we propose one hybrid (SCAN0) and three double-hybrid (SCAN0-DH, SCAN-QIDH, and SCAN0-2) density functionals, which are free from any fitted parameters. The SCAN-based 
double-hybrid functionals consistently outperform their parent SCAN semilocal functional for self-interaction problems and noncovalent interactions. In particular, SCAN0-2, which includes about $79\%$ of 
Hartree-Fock exchange and $50\%$ of second-order M\o ller-Plesset correlation, is shown to be reliably accurate for a very diverse range of applications, such as thermochemistry, kinetics, noncovalent 
interactions, and self-interaction problems. 

\end{abstract} 

\maketitle

\section{Introduction} 

Since the 1990s, Kohn-Sham density functional theory (KS-DFT) \cite{HK,KS} has emerged as one of the most popular methods for studying the ground-state properties of large electronic systems, due to its low 
computational complexity and reasonable accuracy. However, as the exact exchange-correlation (XC) energy functional $E_{xc}[\rho]$ in KS-DFT has not been found, density functional approximations (DFAs) 
for $E_{xc}[\rho]$ have been constantly developed to improve the accuracy of KS-DFT for a diverse range of applications \cite{Parr,DFTreview,DFTreview2,DFTreview3}. 

The various levels of DFAs to $E_{xc}[\rho]$ have been commonly categorized into five different rungs of Jacob's ladder, connecting the Hartree approximation on earth to the heaven of chemical 
accuracy \cite{ladder}. The first rung of the ladder is the local density approximation (LDA) \cite{LDAX,LDAC}, which represents the XC energy density by the local electron density. By construction, LDA is exact 
for a uniform electron gas, which provides a good starting point for more accurate and sophisticated DFAs. Going beyond the LDA, the second rung of the ladder is the generalized gradient approximation (GGA), 
modeling the XC energy density by the local electron density and its gradient to capture the effects of inhomogeneities. The third rung of the ladder is the meta-GGA (MGGA), which incorporates the ingredients 
of the noninteracting Kohn-Sham (KS) kinetic energy density and/or the Laplacian of the electron density into the GGA. 

The functionals on the first three rungs are often called semilocal density functionals. They are reliably accurate for the properties governed by short-range XC effects, and are computationally efficient for large 
systems. Besides, as more exact constraints are likely to be satisfied by construction with the functionals on higher rungs of the ladder, semilocal density functionals are expected to achieve higher accuracy by 
climbing up the ladder at the expense of a slight increase in computational cost \cite{ladder}. However, due to the lack of an accurate description of nonlocal XC effects, semilocal density functionals can yield 
erroneous results in situations where the self-interaction error (SIE), noncovalent interaction error (NCIE), or static correlation error (SCE) is 
enormous \cite{Parr,DFTreview,DFTreview2,DFTreview3,Dobson,SLR-vdW,SciYang,TAO-DFT,TAO-DFT2,TAO-DFT3}. 

Recently, there has been an increasing interest in the nonempirical semilocal density functionals developed by Perdew and co-workers, such as the PBE GGA functional \cite{PBE} and the TPSS MGGA 
functional \cite{TPSS}, demonstrating the usefulness of the functionals developed by the method of constraint satisfaction (without fitting to a large set of experimental or high-level theoretical data) \cite{ladder}. 
Very recently, Sun, Ruzsinszky, and Perdew proposed the strongly constrained and appropriately normed (SCAN) functional \cite{SCAN}, a nonempirical MGGA satisfying all 17 known exact constraints that a 
semilocal functional can \cite{SCAN2}. Besides, SCAN is exact or nearly exact for a set of appropriate norms, including rare-gas atoms and noncovalent interactions. Nevertheless, as SCAN is a semilocal 
functional, it can yield qualitative failures for the SIE, NCIE, and SCE problems. 

Aiming to reduce the SIE and NCIE associated with the SCAN semilocal functional, in this work, we propose SCAN-based hybrid and double-hybrid density functionals based on four existing hybrid and 
double-hybrid models without fitted parameters. The rest of this paper is organized as follows. We describe the models without fitted parameters for the SCAN-based hybrid and double-hybrid functionals in 
Section II, and computational details in Section III. In Section IV, we examine the performance of SCAN-based and PBE-based semilocal, hybrid, and double-hybrid functionals for a diverse range of applications. 
Our conclusions are given in Section V.

\section{Hybrid and Double-Hybrid Models without Fitted Parameters} 

\subsection{DFA0 Hybrid Model} 

In KS-DFT, the adiabatic-connection (AC) formalism \cite{AC1,AC2,AC3,AC4,AC5} provides a very powerful approach to the development of accurate and theoretically justifiable $E_{xc}[\rho]$. Based on the 
AC formalism, the electron-electron interaction ($\alpha \hat{V}_{ee}$) of a system is switched on by a linear scaling using a coupling-strength parameter $\alpha$. Accordingly, the system changes from the 
noninteracting KS reference system ($\alpha=0$) to the fully interacting real system ($\alpha=1$), through a continuum of partially interacting systems ($0\le\alpha\le1$), all of which have the same ground-state 
electron density $\rho({\bf r})$ as that of the fully interacting real system. Consequently, $E_{xc}[\rho]$ can be formally expressed as an integral over $\alpha$: 
\begin{equation} 
E_{xc}[\rho] = \int_{0}^{1} E_{xc,\alpha} d\alpha, 
\label{eq:AC1} 
\end{equation} 
where the AC integrand 
\begin{equation} 
E_{xc,\alpha} = \langle \Psi_{\alpha} \mid \hat{V}_{ee} \mid \Psi_{\alpha} \rangle - \frac{e^2}{2} \int\int \frac{\rho({\bf r})\rho({\bf r'})}{|{\bf r} - {\bf r'}|}d{\bf r}d{\bf r'} 
\label{eq:AC2} 
\end{equation} 
is the potential energy of exchange-correlation at intermediate coupling strength $\alpha$, with $\Psi_{\alpha}$ being the corresponding ground-state wavefunction. 

While the exact $E_{xc,\alpha}$ remains unknown, the noninteracting limit ($\alpha=0$) is given by the Hartree-Fock (HF) exchange energy of the KS orbitals (due to the lack of electron correlation at 
$\alpha=0$): 
\begin{equation} 
E_{xc,\alpha=0} = E_{x}^{\text{HF}}, 
\label{eq:AC3} 
\end{equation} 
and the fully interacting limit ($\alpha=1$) can be well approximated by that of a DFA semilocal functional: 
\begin{equation} 
E_{xc,\alpha=1} \approx E_{xc,\alpha=1}^{\text{DFA}}, 
\label{eq:AC4} 
\end{equation} 
as the XC hole is deeper and thus more localized around its electron at $\alpha=1$ than at $\alpha=0$ \cite{hybrid1,hybrid2,DFA0,DFA0a}. 
Note that $E_{xc,\alpha}^{\text{DFA}}$ (i.e., a DFA to $E_{xc,\alpha}$) can be obtained from a DFA semilocal functional $E_{xc}^{\text{DFA}}[\rho]$ via the coordinate scaling \cite{LP85,LYP85}: 
\begin{equation} 
E_{xc,\alpha}^{\text{DFA}} = \frac{\partial}{\partial \alpha} \{\alpha^{2} E_{xc}^{\text{DFA}}[\rho_{1/\alpha}]\}, 
\label{eq:CS} 
\end{equation} 
where $\rho_{1/\alpha}({\bf r}) = \alpha^{-3} \rho({\bf r}/\alpha)$ is the coordinate-scaled electron density. 

Therefore, a hybrid functional (the fourth rung of the ladder) \cite{hybrid1,hybrid2,B3LYP,DFA0,DFA0a,PBE0}, which incorporates a fraction of HF exchange into a DFA semilocal functional, can be 
theoretically justified by the AC formalism using a simple integrand $E_{xc,\alpha}$ that connects the noninteracting [Eq.\ (\ref{eq:AC3})] and fully interacting [Eq.\ (\ref{eq:AC4})] limits. Due to the adoption of 
an improved $E_{xc,\alpha}$, a hybrid functional is expected to enhance the description of nonlocal exchange effects and the overall accuracy with respect to its parent DFA semilocal functional, wherein a 
less accurate integrand $E_{xc,\alpha} \approx E_{xc,\alpha}^{\text{DFA}}$ is effectively employed. 

Based on the AC formalism, Perdew, Ernzerhof, and Burke proposed the following DFA0 integrand (where $n \ge 1$ is an integer) \cite{DFA0}: 
\begin{equation} 
E_{xc,\alpha}^{\text{DFA0}} = E_{xc,\alpha}^{\text{DFA}} + (E_{x}^{\text{HF}} - E_{x}^{\text{DFA}})(1 - \alpha)^{n-1}. 
\label{eq:AC5} 
\end{equation} 
As shown in Eq.\ (\ref{eq:AC5}), $E_{xc,\alpha=0}^{\text{DFA0}} = E_{x}^{\text{HF}}$ and $E_{xc,\alpha=1}^{\text{DFA0}} = E_{xc,\alpha=1}^{\text{DFA}}$ can be correctly achieved. 
Substituting Eq.\ (\ref{eq:AC5}) into Eq.\ (\ref{eq:AC1}) yields the DFA0 hybrid model: 
\begin{equation} 
\begin{split} 
E_{xc}^{\text{DFA0}} 
=&\; E_{xc}^{\text{DFA}} + \frac{1}{n} (E_{x}^{\text{HF}} - E_{x}^{\text{DFA}}) \\ 
=&\; E_{x}^{\text{DFA}} + E_{c}^{\text{DFA}} + \frac{1}{n} (E_{x}^{\text{HF}} - E_{x}^{\text{DFA}}) \\ 
=&\; \frac{1}{n} E_{x}^{\text{HF}} + (1 - \frac{1}{n}) E_{x}^{\text{DFA}} + E_{c}^{\text{DFA}}, 
\end{split}
\label{eq:DFA0} 
\end{equation} 
where $E_{x}^{\text{HF}}$ is the HF exchange energy, $E_{x}^{\text{DFA}}$ is the DFA exchange energy, $E_{c}^{\text{DFA}}$ is the DFA correlation energy, and $n = 4$ is chosen due to the generally good 
performance of the fourth-order M\o ller-Plesset perturbation theory (MP4) for most molecules \cite{DFA0}. 

To construct a DFA0 hybrid functional without fitted parameters, it is essential to adopt a nonempirical DFA semilocal functional in Eq.\ (\ref{eq:DFA0}). By adopting PBE as the parent DFA functional, one 
obtains PBE0 \cite{PBE0,PBE0a}, a very popular PBE-based hybrid functional. In this work, we adopt the recently developed SCAN semilocal functional as the underlying DFA in Eq.\ (\ref{eq:DFA0}), and 
denominate the resulting SCAN-based hybrid functional as SCAN0.

\subsection{DFA0-DH, DFA-QIDH, and DFA0-2 Double-Hybrid Models} 

In the DFA0 hybrid model, while the exchange part of a DFA is enhanced with the nonlocal character, the correlation part remains the same. Consequently, a DFA0 hybrid functional can fail to describe the 
properties governed by nonlocal correlation effects, such as noncovalent interactions. To make progress, relevant physical constraints may be imposed on the AC integrand $E_{xc,\alpha}$. 

At the weakly interacting limit $(\alpha \rightarrow 0)$, $E_{xc,\alpha}$ has a perturbation expansion \cite{GL1,GL2,GL3}: 
\begin{equation} 
E_{xc,\alpha} = E_{x}^{\text{HF}} + 2 E_{c}^{\text{GL2}} \alpha + \dots, 
\label{eq:AC6} 
\end{equation} 
where $E_{c}^{\text{GL2}}$ is the second-order G\"orling-Levy (GL2) correlation energy, which can be well approximated by the second-order M\o ller-Plesset (MP2) correlation energy of the KS orbitals 
for most of the systems \cite{GL2_MP2}: 
\begin{equation} 
E_{c}^{\text{GL2}} \approx E_{c}^{\text{MP2}}. 
\label{eq:AC7} 
\end{equation} 
Therefore, a double-hybrid (DH) functional (the fifth rung of the 
ladder) \cite{DH0,DH1,DH2,B2PLYP,XYG3,wB97X-2,B2PLYPD3,DHrigo,PBE0-DH,TS2011,xDH-PBE0,PBE0-2,1DH-TPSS,PBE-ACDH,PBE-QIDH,QACF-2,CIDH,H-QIDH}, combining a fraction of HF exchange 
and a fraction of MP2 correlation with a DFA semilocal functional, can be theoretically justified by the AC formalism employing a simple interpolation between the weakly interacting [Eq.\ (\ref{eq:AC6})] and fully 
interacting [Eq.\ (\ref{eq:AC4})] limits of $E_{xc,\alpha}$ as a function of $\alpha$. Owing to the use of an improved version of $E_{xc,\alpha}$, a double-hybrid functional should improve the description of 
nonlocal XC effects and the overall accuracy with respect to its parent DFA semilocal functional. Note that a double-hybrid functional can be generally expressed as 
\begin{equation} 
E_{xc}^{\text{DH}} = a_{x} E_{x}^{\text{HF}} + (1 - a_{x}) E_{x}^{\text{DFA}} + (1 - a_{c}) E_{c}^{\text{DFA}} + a_{c} E_{c}^{\text{MP2}}, 
\label{eq:DH} 
\end{equation} 
where $E_{c}^{\text{MP2}}$ is the MP2 correlation energy, a perturbative term evaluated with the orbitals obtained using the first three terms. 
The two mixing parameters $\{a_{x},a_{c}\}$ can be determined by empirical fitting or physical arguments. 

Recently, Sharkas {\it et al.} \cite{DHrigo} provided a rigorous theoretical justification for double-hybrid functionals based on the AC formalism, leading to the density-scaled one-parameter double-hybrid (DS1DH) 
approximation. Following the DS1DH approximation, Toulouse {\it et al.} \cite{TS2011} proposed a sensible approximation to the density-scaled correlation functional (see Eq.\ (9) of Ref.\ \cite{TS2011}), yielding 
the linearly scaled one-parameter double-hybrid (LS1DH) approximation, wherein a cubic relation between the two mixing parameters $\{a_{x},a_{c}\}$ in a double-hybrid functional is shown: 
\begin{equation} 
a_{c} = (a_{x})^{3}. 
\label{eq:LS1DH} 
\end{equation} 
After substituting Eq.\ (\ref{eq:LS1DH}) into Eq.\ (\ref{eq:DH}), only one parameter (either $a_{x}$ or $a_{c}$) needs to be determined. 

To determine $\{a_{x},a_{c}\}$ without empirical fitting, various arguments were recently proposed and applied to the LS1DH approximation, yielding the following three double-hybrid models 
(in order of increasing $a_{x}$ or $a_{c}$): 
\begin{itemize} 
\item DFA0-DH model \cite{PBE0-DH}: $\{a_{x} = 1/2, \ \ a_{c} = (1/2)^{3} = 1/8\}$ 
\item DFA-QIDH model \cite{PBE-QIDH}: $\{a_{x} = (1/3)^{1/3} \approx 0.693361, \ \ a_{c} = 1/3\}$ 
\item DFA0-2 model \cite{PBE0-2}: $\{a_{x} = (1/2)^{1/3} \approx 0.793701, \ \ a_{c} = 1/2\}$ 
\end{itemize} 

To develop a double-hybrid functional from one of these double-hybrid models without fitted parameters, a nonempirical DFA semilocal functional should be employed in Eq.\ (\ref{eq:DH}). In previous studies, 
PBE was commonly adopted as the underlying DFA in these double-hybrid models, and the corresponding double-hybrid functionals were denominated PBE0-DH \cite{PBE0-DH} for the DFA0-DH model, 
PBE-QIDH \cite{PBE-QIDH} for the DFA-QIDH model, and PBE0-2 \cite{PBE0-2} for the DFA0-2 model. In this work, we adopt SCAN as the parent DFA semilocal functional in the DFA0-DH, DFA-QIDH, and 
DFA0-2 models, and denominate the resulting SCAN-based double-hybrid functionals as SCAN0-DH, SCAN-QIDH, and SCAN0-2, respectively.

\section{Computational Details} 

For a comprehensive comparison, we examine the following 10 PBE-based and SCAN-based semilocal, hybrid, and double-hybrid functionals: 
\begin{itemize} 
\item DFA semilocal functionals: PBE \cite{PBE} and SCAN \cite{SCAN} 
\item DFA0 hybrid functionals: PBE0 \cite{PBE0,PBE0a} and SCAN0 
\item DFA0-DH double-hybrid functionals: PBE0-DH \cite{PBE0-DH} and SCAN0-DH 
\item DFA-QIDH double-hybrid functionals: PBE-QIDH \cite{PBE-QIDH} and SCAN-QIDH 
\item DFA0-2 double-hybrid functionals: PBE0-2 \cite{PBE0-2} and SCAN0-2 
\end{itemize} 
on various test sets involving 
\begin{itemize} 
\item the 223 atomization energies (AEs) of the G3/99 set \cite{G399a,G399b,G399c} 
\item the 40 ionization potentials (IPs), 25 electron affinities (EAs), and 8 proton affinities (PAs) of the G2-1 set \cite{G21} 
\item the 76 barrier heights of the NHTBH38/04 and HTBH38/04 sets \cite{ZL2004_2,ZG2005} 
\item the 22 noncovalent interactions of the S22 set \cite{S22,S22a} 
\item the 66 noncovalent interactions of the S66 set \cite{S66} 
\item two interaction energy curves of the benzene dimer from the S66$\times8$ set \cite{S66} 
\item seven isodesmic reaction energies of $n$-alkanes to ethane \cite{RZ2000,SE1,ST2010,G2010,SE2} 
\item two dissociation energy curves of symmetric radical cations \cite{SIE1,SIE2,SIE3,SIE4,SIE5} 
\item the dissociation energy curve of $\text{H}_{2}$ 
\end{itemize} 
Detailed information about these test sets can be found in Refs.\ \cite{wB97,PBE0-2,TAO-DFT}. 

All calculations are performed with a development version of \textsf{Q-Chem 4.3} \cite{Q-Chem}. Spin-restricted theory is used for singlet states and spin-unrestricted theory for others, unless noted otherwise. 
Results are computed using the 6-311++G(3df,3pd) basis set with the ultrafine grid EML(99,590), consisting of 99 Euler-Maclaurin radial grid points \cite{EM} and 590 Lebedev angular grid points \cite{L}. 
The MP2 calculations are done without frozen core (i.e., all electrons are included in the perturbation). 
For computational efficiency, the resolution-of-identity (RI) approximation \cite{KF1997} is used for calculations with the MP2 correlation (using sufficiently large auxiliary basis sets). 
For the interaction energies of the weakly bound systems, the counterpoise correction \cite{CP} is employed to reduce basis set superposition error (BSSE). 
The error for each entry is defined as (error = theoretical value $-$ reference value). The notation used for characterizing statistical errors is as follows: mean signed errors (MSEs), mean absolute errors (MAEs), 
root-mean-square (rms) errors, maximum negative errors (Max($-$)), and maximum positive errors (Max(+)).

\section{Results and Discussion} 

For the test sets shown in \Cref{table:testsets}, the functionals on higher rungs of Jacob's ladder generally perform better than those on lower rungs, though there are some exceptions \cite{supp}. For the AEs 
of the G3/99 set, the SCAN0-2 double-hybrid functional has the best performance. Surprisingly, the SCAN semilocal functional performs comparably to the PBE0-DH, PBE-QIDH, PBE0-2, and SCAN-QIDH 
double-hybrid functionals and the PBE0 hybrid functional. SCAN and SCAN0-2 are consistently better than their PBE-based counterparts (i.e., PBE and PBE0-2, respectively) in performance, while SCAN0, 
SCAN0-DH, and SCAN-QIDH fail to outperform their PBE-based counterparts (i.e., PBE0, PBE0-DH, and PBE-QIDH, respectively). Nevertheless, for the IPs, EAs, and PAs of the G2-1 set, all the functionals 
have similar performance. 

For the barrier heights of the NHTBH38/04 and HTBH38/04 sets, SCAN slightly outperforms PBE, whereas SCAN0 and PBE0 are significantly better than SCAN and PBE, respectively, in performance. 
PBE-QIDH performs best, followed by SCAN-QIDH. Overall, the SCAN-based and PBE-based double-hybrid functionals are comparable in performance, and are much more accurate than the corresponding 
hybrid and semilocal functionals. 

For the noncovalent interactions of the S22 and S66 sets, the SCAN-based semilocal, hybrid, and double-hybrid functionals consistently outperform their PBE-based counterparts, which may be attributed to 
the improved performance of SCAN for noncovalent interactions (mainly due to its improved treatment of medium-range dynamical correlation effects), with respect to PBE. Nonetheless, as SCAN and PBE 
are both semilocal functionals, they cannot adequately describe the van der Waals (vdW) asymptote \cite{Dobson,SLR-vdW}, which requires an accurate treatment of long-range dynamical correlation effects. 
Incorporating SCAN and PBE with the double-hybrid 
schemes \cite{DH0,DH1,DH2,B2PLYP,XYG3,wB97X-2,B2PLYPD3,DHrigo,PBE0-DH,TS2011,xDH-PBE0,PBE0-2,1DH-TPSS,PBE-ACDH,PBE-QIDH,QACF-2,CIDH,H-QIDH}, 
the DFT-D (KS-DFT with empirical dispersion corrections) schemes \cite{DFT-D1,DFT-D2,wB97X-D,B2PLYPD3,Sherrill,DFT-D3appl,wM05-D,LC-D3}, or fully nonlocal density 
functionals \cite{vdW,VV10,VV10test} may greatly improve the accuracy of SCAN and PBE for vdW interactions. Among the functionals examined on the S22 and S66 sets, SCAN0-2 ranks first, while 
SCAN-QIDH and PBE0-2 rank second and third, respectively. SCAN0-DH only slightly improves upon SCAN due to its relatively small fraction (12.5\%) of MP2 correlation. To the best of our knowledge, the 
MAEs of SCAN0-2 on the S22 and S66 sets are record low. Therefore, SCAN0-2 should outperform most existing density functionals here, and possibly, for other noncovalent interactions as well. 

\begin{figure} 
\includegraphics[scale=1.3]{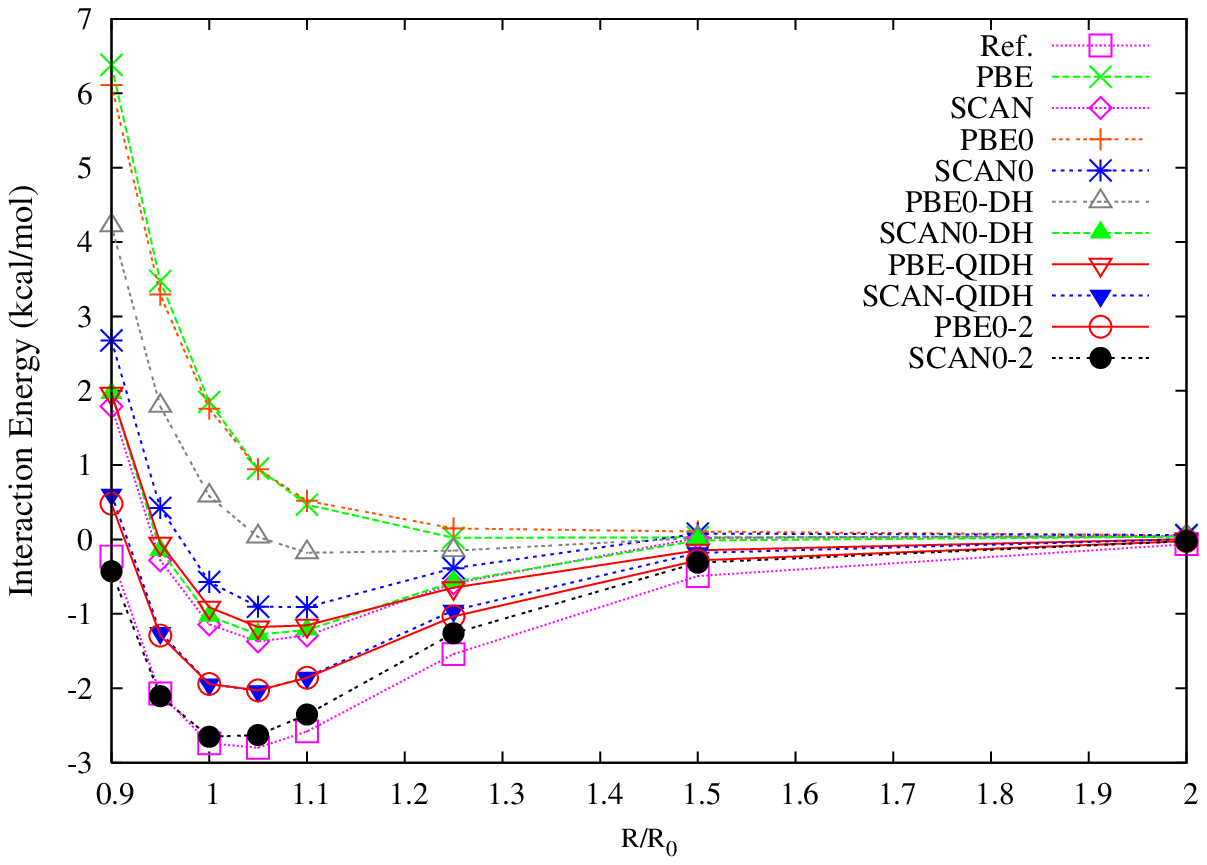} 
\caption{\label{fig:S66x8_BBpp} 
Interaction energy curve for the parallel-displaced configuration of the benzene dimer as a function of the intermonomer distance $R$ (defined in the S66$\times$8 set \cite{S66}), 
where $R_0$ is the equilibrium distance.} 
\end{figure}

\begin{figure} 
\includegraphics[scale=1.3]{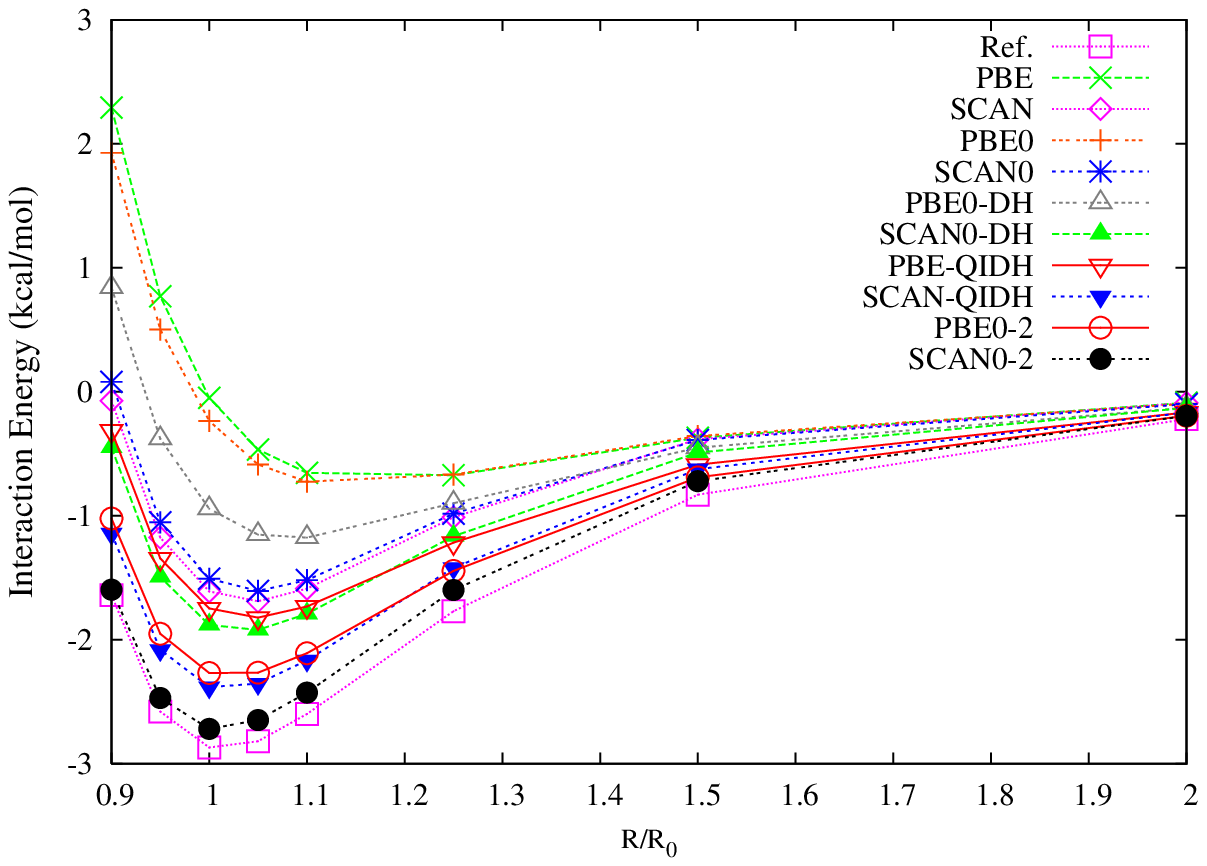} 
\caption{\label{fig:S66x8_BBTS} 
Interaction energy curve for the T-shaped configuration of the benzene dimer as a function of the intermonomer distance $R$ (defined in the S66$\times$8 set \cite{S66}), 
where $R_0$ is the equilibrium distance.} 
\end{figure}

In addition, our investigation is extended to the interaction energy curves of the benzene dimer in the S66$\times8$ set. Specifically, we calculate the interaction energy curves for the parallel-displaced 
(in \Cref{fig:S66x8_BBpp}) and T-shaped (in \Cref{fig:S66x8_BBTS}) configurations of the benzene dimer as functions of the intermonomer distance $R$ (defined in Ref.\ \cite{S66}), where the optimized 
geometries and reference values are taken from the S66$\times8$ set. Overall, the predicted interaction energy curves of SCAN0-2 are extremely close to the accurate reference curves (within an error of 
0.28 kcal/mol), followed by PBE0-2 (within an error of 0.80 kcal/mol) and SCAN-QIDH (within an error of 0.84 kcal/mol). All the other functionals predict severely underbinding or even repulsive curves. 

\begin{figure} 
\includegraphics[scale=1.3]{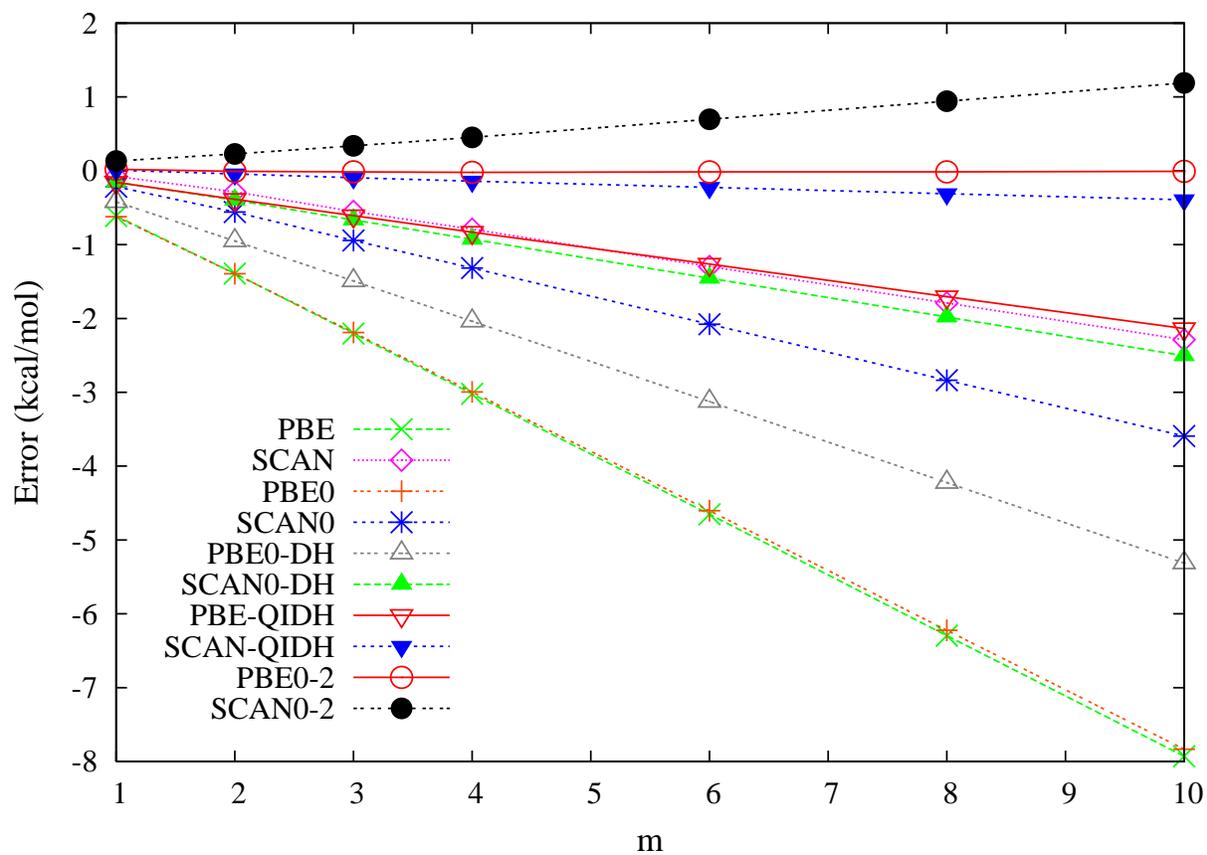} 
\caption{\label{fig:alkanes} 
Errors for isodesmic reaction energies of $n$-alkanes to ethane.} 
\end{figure}

The isodesmic reaction energies (see Eq.\ (\ref{eq:alkanes}) for the considered bond separation reaction) of the linear conformations of the {\it n}-alkanes to form ethane have been discovered to yield 
systematic errors in standard KS-DFT calculations, when the number of protobranches in {\it n}-alkanes, $m$, increases \cite{RZ2000,SE1,ST2010,G2010,SE2}. 
\begin{equation} 
\text{CH}_{3}(\text{CH}_{2} )_{m}\text{CH}_{3} + m\text{CH}_{4} \rightarrow (m+1)\text{C}_{2}\text{H}_{6}. 
\label{eq:alkanes} 
\end{equation} 
To assess if the SCAN-based hybrid and double-hybrid functionals alleviate this problem, we calculate seven isodesmic reaction energies of {\it n}-alkanes to ethane, where the optimized geometries and 
reference values are taken from Ref.\ \cite{G2010}. As shown in \Cref{fig:alkanes}, due to the accurate treatment of medium-range electron correlation effects \cite{G2010}, the predicted reaction energies 
of PBE0-2 are the closest ones to the reference values (within an error of 0.02 kcal/mol), followed by SCAN-QIDH (within an error of 0.39 kcal/mol) and SCAN0-2 (within an error of 1.19 kcal/mol). By contrast, 
all the other functionals exhibit considerable errors with the increase of $m$. 

\begin{figure} 
\includegraphics[scale=1.3]{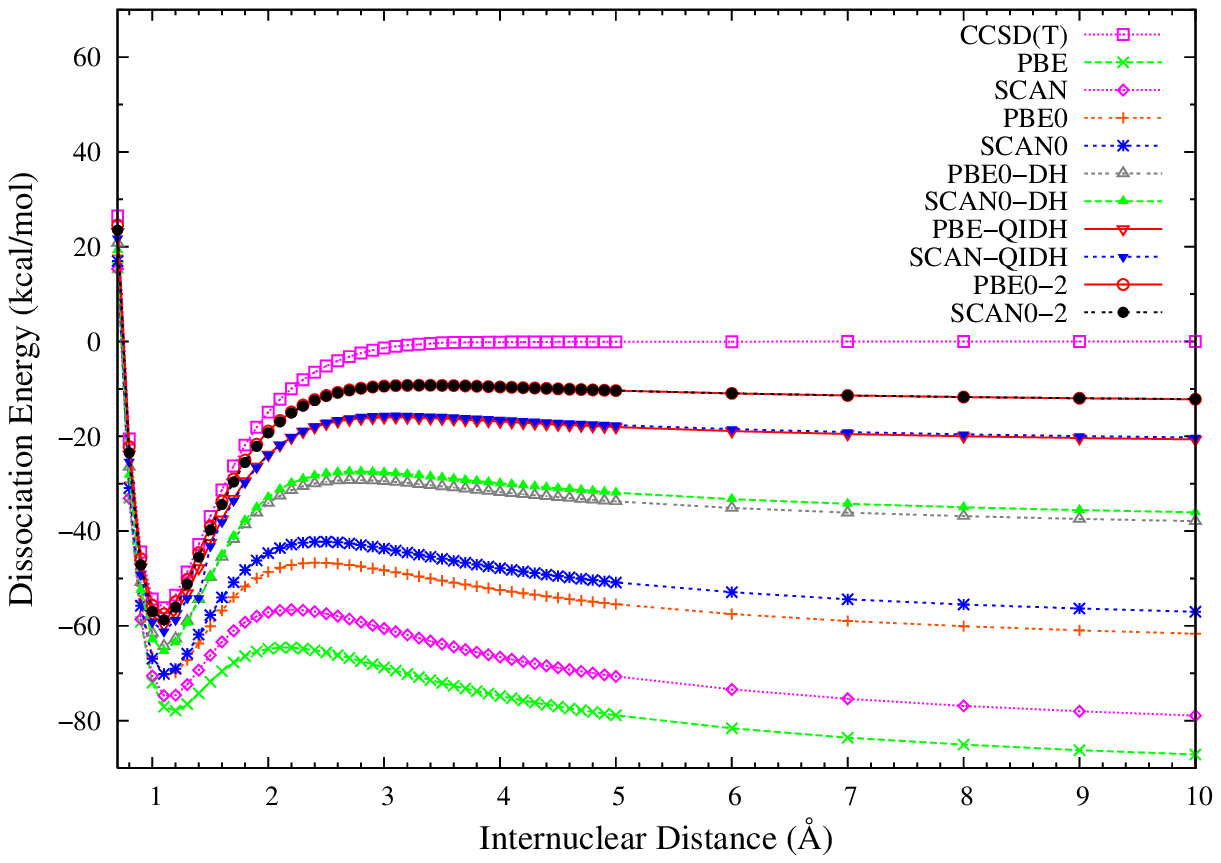} 
\caption{\label{fig:he2p} 
Dissociation energy curve of $\text{He}_{2}^{+}$. Zero level is set to \textit{E}(He) + \textit{E}($\text{He}^{+}$) for each method.} 
\end{figure}

\begin{figure} 
\includegraphics[scale=1.3]{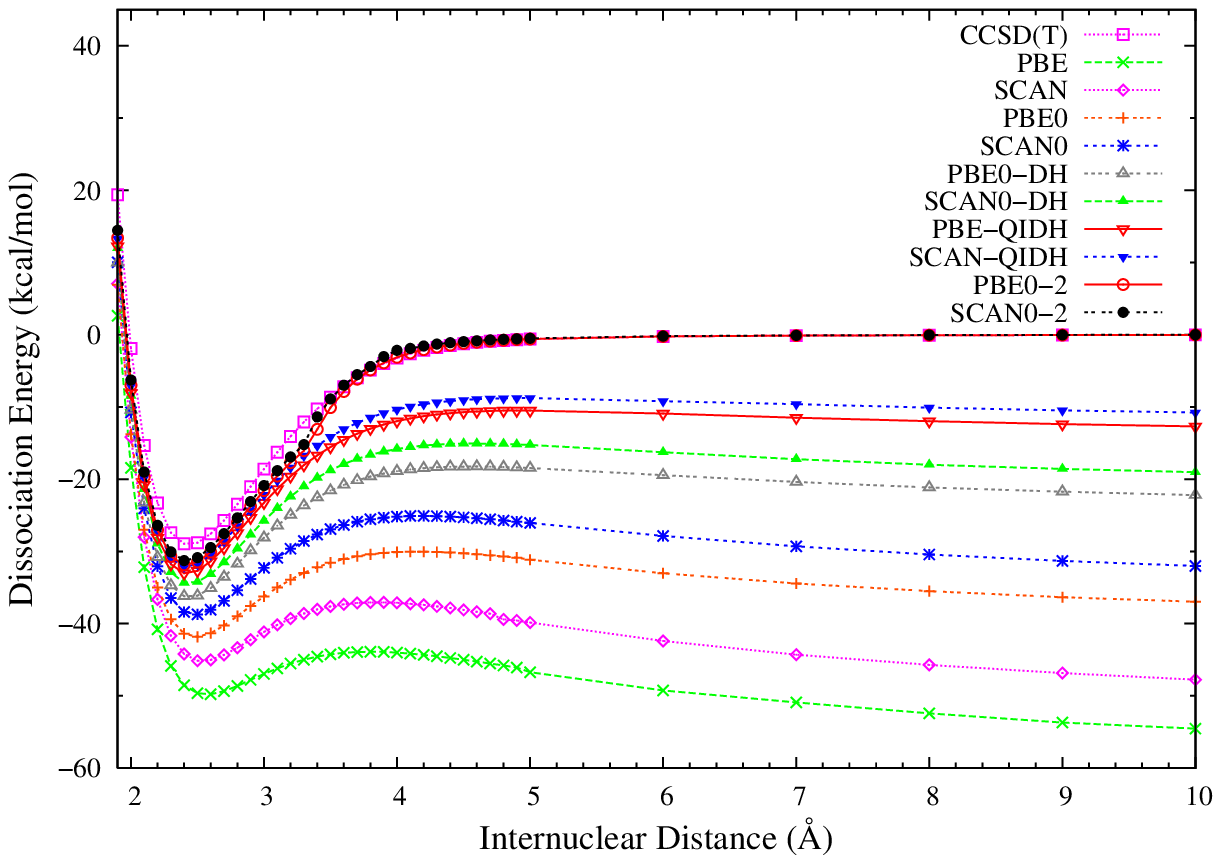} 
\caption{\label{fig:ar2p} 
Dissociation energy curve of $\text{Ar}_{2}^{+}$. Zero level is set to \textit{E}(Ar) + \textit{E}($\text{Ar}^{+}$) for each method.} 
\end{figure}

Owing to the severe SIEs associated with semilocal functionals, spurious fractional charge dissociation can occur for symmetric charged radicals \cite{SIE1,SIE2,SIE3,SIE4,SIE5}. To investigate how the 
SCAN-based hybrid and double-hybrid functionals improve upon the SIE problems, spin-unrestricted calculations are performed for the dissociation energy curves of $\text{He}_{2}^{+}$ and $\text{Ar}_{2}^{+}$. 
The results are compared with those calculated using the highly accurate CCSD(T) theory \cite{CCSD(T)}. As shown in \Cref{fig:he2p}, the predicted $\text{He}_{2}^{+}$ dissociation energy curves of 
SCAN0-2 and PBE0-2 are very close to the CCSD(T) curve. It appears that the SIEs associated with SCAN0-2 and PBE0-2 are more than six times smaller than their parent semilocal functionals, SCAN and 
PBE, respectively. As can be seen in \Cref{fig:ar2p}, in contrast to the other functionals, both SCAN0-2 and PBE0-2 can dissociate $\text{Ar}_{2}^{+}$ correctly, which are indeed very promising. 
Note that a discontinuity undesirably appears in the derivative of the SCAN0-2 (or PBE0-2) dissociation energy curve for $\text{Ar}_2^{+}$, due to the use of MP2 correlation (as discussed in 
Refs.\ \cite{N-rep,wB97X-2,PBE0-2}). Overall, the SCAN-based semilocal, hybrid, and double-hybrid functionals consistently outperform their PBE-based counterparts. 

\begin{figure} 
\includegraphics[scale=1.3]{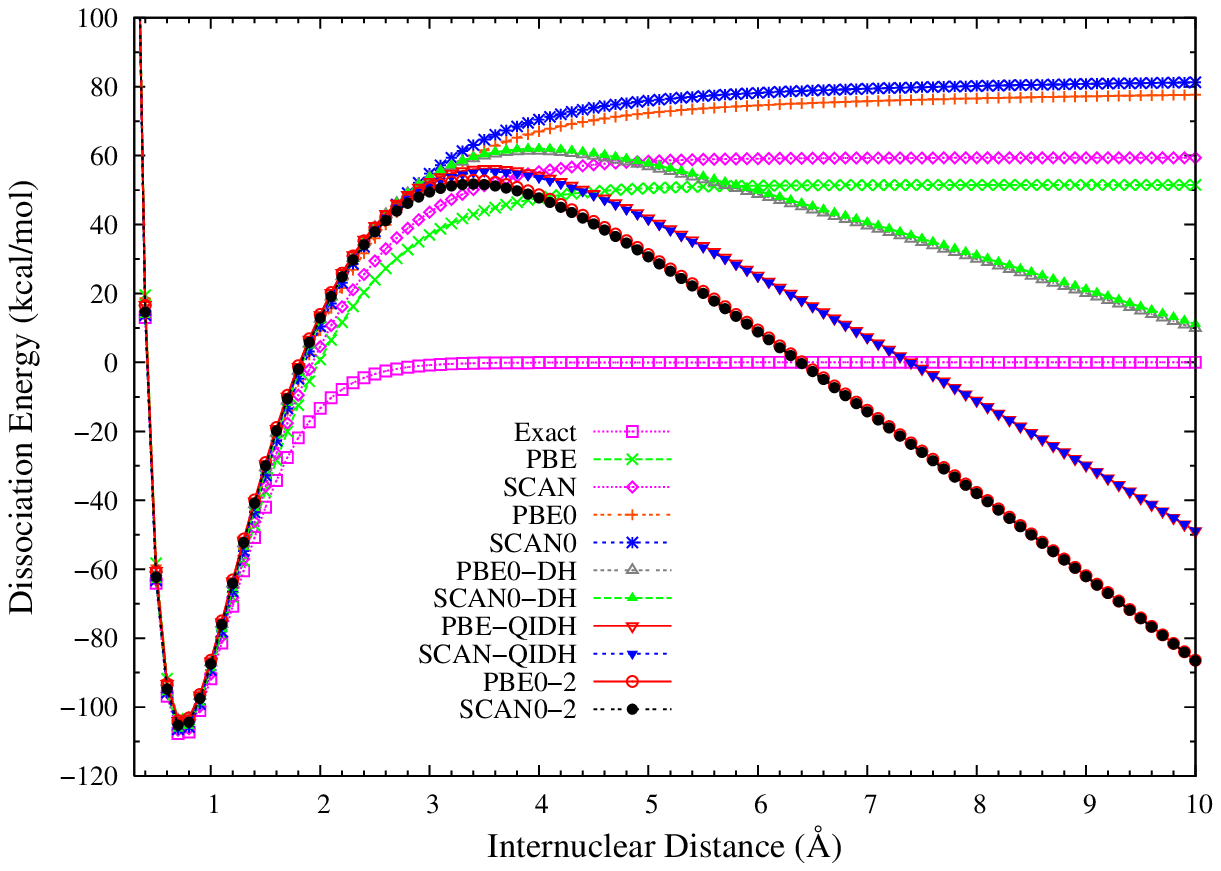} 
\caption{\label{fig:h2} 
Spin-restricted dissociation energy curve of $\text{H}_{2}$. The exact curve is calculated using the CCSD theory. Zero level is set to 2\textit{E}(H) for each method.} 
\end{figure}

Due to the presence of strong static correlation effects, $\text{H}_{2}$ dissociation (a single-bond breaking system) has been very challenging for conventional density functionals in KS-DFT. Based on the 
symmetry constraint, the spin-restricted and spin-unrestricted dissociation energy curves of $\text{H}_{2}$ calculated by the exact theory, should be identical, implying that $\text{H}_{2}$ should be properly 
separated into two $\text{H}$ atoms at the dissociation limit. Therefore, the difference between the dissociation limit of the spin-restricted dissociation energy curve and the sum of the energies of two 
$\text{H}$ atoms, can be adopted as a quantitative measure of SCEs of approximate density functional methods \cite{SciYang,TAO-DFT,TAO-DFT2}. To examine the performance of the SCAN-based and 
PBE-based functionals upon the SCE problems, spin-restricted calculations are performed for the dissociation energy curves of $\text{H}_{2}$. As shown in \Cref{fig:h2}, all the functionals perform comparably 
to the CCSD theory \cite{CCSD} (exact for any two-electron system) near the equilibrium bond length of $\text{H}_{2}$, where the single-reference character is predominant. Nevertheless, all the functionals 
yield enormous SCEs at the dissociation limit, where the multi-reference character becomes significant. It is worthwhile to note that any double-hybrid functional must fail for $\text{H}_{2}$ dissociation, as the 
MP2 correlation energy unphysically diverges to minus infinity for systems with vanishing energy differences between the highest occupied and lowest unoccupied molecular orbitals (HOMO-LUMO gaps). 
Note also that hybrid and double-hybrid functionals may not improve upon the parent semilocal functionals for the SCE problems \cite{SciYang,TAO-DFT,TAO-DFT2}.

\section{Conclusions} 

In summary, by incorporating the nonempirical SCAN semilocal density functional as the parent DFA functional in the hybrid and double-hybrid models discussed in this work, we have developed one hybrid 
(SCAN0) and three double-hybrid (SCAN0-DH, SCAN-QIDH, and SCAN0-2) density functionals, which are free from any fitted parameters. Owing to the significant improvement over their parent semilocal 
functional SCAN (the third rung of Jacob's ladder) for a wide range of systems, SCAN0 fits reasonably well into the fourth rung of the ladder, while SCAN0-DH, SCAN-QIDH, and SCAN0-2 fit well into the 
fifth rung of the ladder. For the SIE and NCIE problems, the SCAN-based semilocal, hybrid, and double-hybrid functionals consistently outperform their PBE-based counterparts. In particular, SCAN0-2 is 
generally superior in performance for a very diverse range of applications (including the SIE and NCIE problems, but not the SCE problems), extending the applicability of SCAN-based functionals to a very 
wide variety of systems. It remains very difficult to devise a generally accurate density functional method resolving the SIE, NCIE, and SCE problems at affordable computational costs. Work in this direction 
is in progress.

\begin{acknowledgments} 

This work was supported by the Ministry of Science and Technology of Taiwan (Grant No.\ MOST104-2628-M-002-011-MY3), National Taiwan University (Grant No.\ NTU-CDP-104R7818), 
the Center for Quantum Science and Engineering at NTU (Subproject Nos.:\ NTU-ERP-104R891401 and NTU-ERP-104R891403), and the National Center for Theoretical Sciences of Taiwan. 

\end{acknowledgments}

\bibliographystyle{jcp}

\newpage 

\newpage 

\newpage 

\newpage 

\newpage 

\newpage 

\newpage 
\begin{table*} 
\scriptsize 
\caption{\label{table:testsets} 
Statistical errors (in kcal/mol) of the PBE-based and SCAN-based semilocal, hybrid, and double-hybrid density functionals for the 223 atomization energies (AEs) of the G3/99 set \cite{G399a,G399b,G399c}, 
the 40 ionization potentials (IPs), 25 electron affinities (EAs), and 8 proton affinities (PAs) of the G2-1 set \cite{G21}, the 76 barrier heights of the NHTBH38/04 and HTBH38/04 sets \cite{ZL2004_2,ZG2005}, 
the 22 noncovalent interactions of the S22 set \cite{S22,S22a}, and the 66 noncovalent interactions of the S66 set \cite{S66}.} 
\begin{ruledtabular} 
\begin{tabular}{lccccccccccc} 
& & \multicolumn{2}{c}{DFA} & \multicolumn{2}{c}{DFA0} & \multicolumn{2}{c}{DFA0-DH} & \multicolumn{2}{c}{DFA-QIDH} & \multicolumn{2}{c}{DFA0-2}  \\ 
\cline{3-4} 
\cline{5-6} 
\cline{7-8} 
\cline{9-10} 
\cline{11-12} 
System & Error & PBE & SCAN & PBE0 & SCAN0 & PBE0-DH & SCAN0-DH & PBE-QIDH & SCAN-QIDH & PBE0-2 & SCAN0-2  \\ 
\hline 
G3/99  & MSE  & 20.90 & 4.27 & 3.91 & -10.18 & 0.24 & -10.52 & 1.54 & -5.96 & 3.08 & -2.18  \\
(223)   & MAE  & 21.51 & 5.52 & 6.30 & 10.50 & 5.11 & 10.77 & 5.37 & 6.31 & 5.82 & 4.18  \\ 
            & rms   & 26.30 & 6.68 & 8.65 & 12.34 & 7.05 & 13.01 & 7.23 & 8.22 & 7.95 & 5.39 \\ 
\hline 
IP     & MSE & 0.04 & -0.08 & 0.19 & -0.05 & 0.22 & -0.39 & -0.27 & -0.29 & -0.32 & -0.44  \\ 
(40)  & MAE & 3.44 & 4.24 & 3.48 & 4.57 & 3.17 & 3.75 & 2.43 & 3.22 & 2.38 & 2.83  \\ 
         & rms  & 4.35 & 5.27 & 4.21 & 5.81 & 3.96 & 4.75 & 2.93 & 3.95 & 3.17 & 3.50  \\ 
\hline 
EA   & MSE & 1.72 & -0.31 & -1.07 & -2.19 & -2.03 & -2.53 & -2.23 & -2.37 & -1.95 & -2.16  \\ 
(25) & MAE & 2.42 & 3.91 & 3.10 & 5.08 & 3.54 & 4.76 & 3.26 & 3.97 & 2.91 & 3.38  \\ 
        & rms  & 3.06 & 4.57 & 3.53 & 6.03 & 4.17 & 5.65 & 3.89 & 4.62 & 3.65 & 4.05  \\ 
\hline 
PA & MSE & -0.83 & -0.03 & 0.18 & 0.01 & 0.47 & -0.25 & 0.33 & -0.52 & 0.07 & -0.67  \\ 
(8) & MAE & 1.60  & 1.17 & 1.14 & 1.27 & 1.09 & 1.32 & 1.04 & 1.20 & 0.96 & 1.07  \\ 
      & rms  & 1.91 & 1.61 & 1.61 & 1.82 & 1.67 & 1.73 & 1.45 & 1.49 & 1.21 & 1.33  \\ 
\hline 
NHTBH & MSE & -8.52 & -7.48 & -3.13 & -3.28 & -0.32 & -0.83 & 1.39 & 0.87 & 2.34 & 1.94  \\ 
(38)       & MAE & 8.62  & 7.62 & 3.63 & 3.84 & 1.57 & 2.24 & 1.62 & 1.63 & 2.44 & 2.03  \\ 
              & rms  & 10.61 & 8.72 & 4.63 & 4.63 & 2.19 & 2.82 & 2.58 & 2.55 & 3.71 & 3.43  \\ 
\hline 
HTBH & MSE & -9.67 & -7.49 & -4.60 & -3.99 & -1.87 & -1.81 & -0.28 & -0.56 & 0.50 & 0.19  \\ 
(38)     & MAE & 9.67  & 7.49 & 4.60 & 4.06 & 1.93 & 2.00 & 0.99 & 1.14 & 1.39 & 1.18  \\ 
            & rms  & 10.37 & 7.94 & 4.88 & 4.46 & 2.19 & 2.48 & 1.30 & 1.43 & 1.74 & 1.45  \\ 
\hline 
S22 & MSE & 2.72 & 0.69 & 2.45 & 0.85 & 1.75 & 0.60 & 1.03 & 0.28 & 0.61 & 0.09  \\ 
(22) & MAE & 2.72 & 0.92 & 2.46 & 1.11 & 1.78 & 0.84 & 1.05 & 0.43 & 0.61 & 0.19  \\ 
        & rms  & 3.73 & 1.22 & 3.45 & 1.54 & 2.49 & 1.19 & 1.42 & 0.59 & 0.78 & 0.25  \\ 
\hline 
S66 & MSE & 2.22 & 0.61 & 2.09 & 0.77 & 1.56 & 0.58 & 1.00 & 0.33 & 0.66 & 0.19  \\ 
(66) & MAE & 2.23 & 0.85 & 2.10 & 1.01 & 1.58 & 0.81 & 1.01 & 0.49 & 0.67 & 0.27  \\ 
        & rms  & 2.75 & 1.04 & 2.61 & 1.25 & 1.98 & 1.01 & 1.27 & 0.63 & 0.83 & 0.40  \\ 
\end{tabular} 
\end{ruledtabular} 
\end{table*} 

\end{document}